\documentclass[twocolumn,prl,showpacs,preprintnumbers,amsmath,amssymb]{revtex4}

\usepackage{graphicx}

\begin{document}


\title{Two- and Three-Dimensional Fermi Surfaces and their Nesting Properties in Superconducting BaFe$_2$(As$_{1-x}$P$_x$)$_2$}

\author{T. Yoshida$^{1,2}$, I. Nishi$^1$, S. Ideta$^1$, A. Fujimori$^{1,2}$, M. Kubota$^3$, K. Ono$^3$, S. Kasahara$^{4,5}$,
 T. Shibauchi$^5$, T. Terashima$^4$, Y. Matsuda$^5$, H. Ikeda$^{2,5}$ and R. Arita$^{2,6}$ }

\affiliation{$^1$Department of Physics, University of Tokyo,
Bunkyo-ku, Tokyo 113-0033, Japan}

\affiliation{$^2$JST, Transformative Research-Project on Iron
Pnictides (TRIP), Chiyoda, Tokyo 102-0075, Japan}

\affiliation{$^3$KEK, Photon Factory, Tsukuba, Ibaraki 305-0801,
Japan}

\affiliation{$^4$Research Center for Low Temperature and Materials
Sciences, Kyoto University, Kyoto 606-8502, Japan}

\affiliation{$^5$ Department of Physics, Kyoto University, Kyoto
606-8502, Japan}

\affiliation{$^6$Department of Applied Physics, University of
Tokyo, Bunkyo-ku, Tokyo 113-8561}
\date{\today}

\begin{abstract}
We have studied the three-dimensional shapes of the Fermi surfaces
(FSs) of BaFe$_2$(As$_{1-x}$P$_x$)$_2$ ($x$=0.38), where
superconductivity is induced by isovalent P substitution, by
angle-resolved photoemission spectroscopy. Moderately strong
electron mass enhancement has been identified for both the
electron and hole FSs. Among two observed hole FSs, the nearly
two-dimensional one shows good nesting with the outer
two-dimensional electron FS, but its orbital character is
different from the outer electron FS. The three-dimensional hole
FS shows poor nesting with the electron FSs. The present results
suggest that the three-dimensionality and the difference in the
orbital character weaken FS nesting, leaving partial nesting among
the outer electron FSs of $d_{xy}$ character and/or that within
the three-dimensional hole FS, which may lead to the nodal
superconductivity.
\end{abstract}


\maketitle

Most of experimental results on the iron-pnictide superconductors
have so far indicated that the superconducting gap opens on the
entire Fermi surfaces \cite{Ding, Terashima}, most likely a
$s\pm$-wave gap, in contrast to the $d$-wave superconducting gap
in the high-$T_c$ cuprate superconductors. However, recent
penetration depth, thermal conductivity \cite{Hashimoto}, and NMR
\cite{Nakai} studies of BaFe$_2$(As$_{1-x}$P$_x$)$_2$
\cite{Kasahara} show signatures of superconducting gap with line
node. In this system, the substitution of P for As suppresses
magnetic order without changing the number of Fe 3$d$ electrons
and induces superconductivity with a maximum $T_c\sim$ 30 K at
$x\sim$0.3. According to theories of spin fluctuation-mediated
superconductivity, line nodes may appear when the pnictogen height
becomes small \cite{Kuroki, Ikeda}, due to changes in nesting
conditions caused by the disappearance of a hole Fermi surface
(FS) of $d_{xy}$ character around the zone center. (Here, the $x$-
and $y$-axis point towards nearest neighbor Fe atoms.)

The importance of FS nesting for the superconductivity has been
pointed out in early studies on Ba$_{1-x}$K$_x$Fe$_2$As$_2$ by
angle-resolved photoemission spectroscopy (ARPES) \cite{Ding,
Terashima} based on models with two-dimensional (2D) electronic
structure. For the family of BaFe$_2$As$_2$ system, however,
strong three-dimensionality in FSs has been identified by the
band-structure calculation \cite{Singh} and confirmed by ARPES
studies \cite{Vilmercati,Malaeb}. Because the P substitution in
BaFe$_2$(As$_{1-x}$P$_x$)$_2$ reduces the $c$-axis length as well
as pnictogen height and increases the inter-layer hopping,
band-structure calculation predicts that the shapes of hole FSs
become more three dimensional with P substitution \cite{Kasahara,
Analytis1, Shishido}. Therefore, it is crucial to reveal the
three-dimensional electronic structure of the
BaFe$_2$(As$_{1-x}$P$_x$)$_2$ superconductor in order to elucidate
the relationship between FS nesting, superconductivity, and gap
symmetry. De Haas-van Alphen (dHvA) measurements \cite{Shishido}
in a wide substitution range (0.41 $< x <$ 1) have indicated a
shrinkage of the electron FSs compared to the band-structure
calculation as one approaches to the optimal composition from the
end material BaFe$_2$P$_2$ ($x$=1). Also, significant electron
mass renormalization has been observed there, which is reminiscent
of heavy Fermion superconductors.

In the present study, we have studied the three-dimensional shapes
of the FSs near optimal composition ($x$= 0.38) by performing
ARPES measurements. Using tunable photon energies of synchrotron
radiation, we have observed FSs predicted by band-structure
calculation in three-dimensional momentum space. Also, we have
observed electron mass renormalization for each FS, quantitatively
consistent with the dHvA results \cite{Shishido}. Based on the
obtained FSs, we shall discuss the FS nesting properties in
three-dimensional momentum space and their implication for the
nodal superconductivity.

High-quality single crystals of BaFe$_2$(As$_{1-x}$P$_x$)$_2$ with
$x$=0.38 ($T_c$=28 K) were grown as described elsewhere
\cite{Kasahara}. ARPES measurements were carried out at BL-28A of
Photon Factory (PF) using circularly-polarized light. A Scienta
SES-2002 analyzer was used with the total energy resolution of
$\sim$ 15 meV and the momentum resolution of $\sim$ 0.02$\pi/a$.
In-plane ($k_X$, $k_Y$) and out-of-plane electron momenta ($k_z$)
are expressed in units of $\pi/a$ and 2$\pi/c$, respectively,
where $a$ = 3.92 \textrm{\AA} and $c$ = 12.8 \textrm{\AA} are the
in-plane and the out-of-plane lattice constants. Here, the
tetragonal unit cell axes are defined as $X$, $Y$, and $z$. The
crystals were cleaved \textit{in situ} at $T$=10 K in an
ultra-high vacuum $\sim$5$\times$10$^{-11}$ Torr. Calibration of
the Fermi level ($E_F$) of the samples was achieved by referring
to that of gold. Our data are compared with a band-structure
calculation performed using a WIEN2K package \cite{wien2k}.

\begin{figure}
\includegraphics[width=7.8cm]{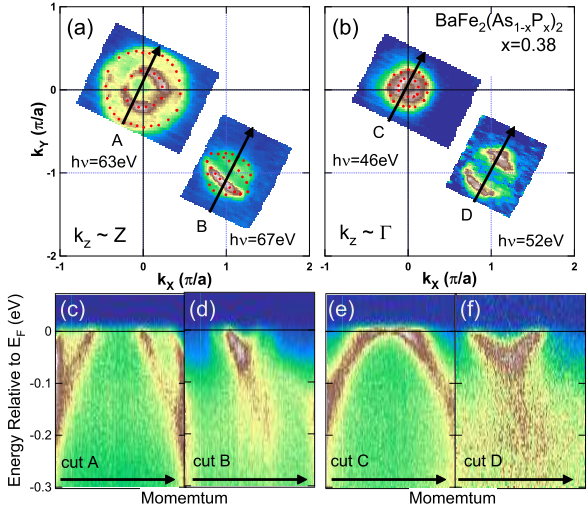}
\caption{\label{Mapping2D}(Color online) Fermi surfaces and band
dispersions of BaFe(As$_{1-x}$P$_x$)$_2$ ($x$=0.38) observed by
ARPES. (a)(b) ARPES intensity at $E_F$ mapped in the $k_X$-$k_Y$
plane taken at several photon energies. Red dots indicate $k_F$
positions determined by the peak positions of momentum
distribution curves (MDC's). (c)-(f) Band dispersion corresponding
to the cuts in panels (a) and (b).}
\end{figure}

FS mapping in the $k_X$-$k_Y$ plane is shown in Figs.
\ref{Mapping2D} (a) and (b). By assuming an inner potential
$V_0$=13.5 eV, panels (a) and (b) approximately correspond to the
$k_X$-$k_Y$ planes including the $Z$ and the $\Gamma$ point,
respectively. We have observed at least two hole FSs around the
center of the 2D Brillouin zone (BZ) and two electron FSs around
the corner of the 2D BZ. One can clearly see a small diameter of
the hole FSs around the $\Gamma$ point compared to those around
the $Z$ point, suggesting strong three-dimensionality of the FSs.
Band dispersions corresponding to cuts in panels (a) and (b) are
illustrated in panels (c)-(f). Particularly, for the electron band
dispersions around the $X$ point, two sheets of FSs have been
observed as shown in panels (a) and (d), consistent with the
band-structure calculation.

\begin{figure}
\includegraphics[width=7.8cm]{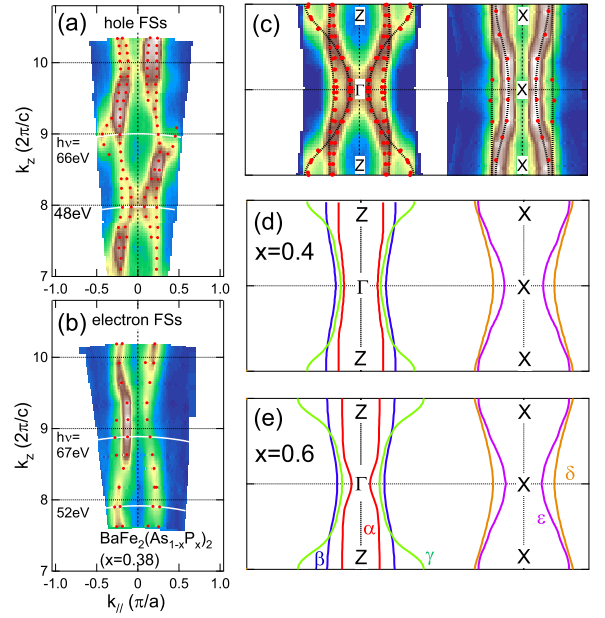}
\caption{\label{Mapping3D}(Color online) Fermi surface mapping in
the $k_{\parallel}$-$k_z$ plane obtained by changing the photon
energy. (a) Hole Fermi surfaces around the center of the 2D
Brillouin zone (BZ). (b) Electron Fermi surfaces around the corner
of the 2D BZ. The directions of $k_{\parallel}$ are shown in Fig.
\ref{Mapping2D}. (c) Fermi surfaces obtained by symmetrizing the
plots in panels (a) and (b). (d)(e) Band-structure calculation for
$x$=0.4 and 0.6.}
\end{figure}

In order to further investigate the three-dimensional electronic
structure, intensity mapping in the $k_{\parallel}$-$k_z$ plane
was performed by changing the photon energy as shown in Fig.
\ref{Mapping3D}. The direction of $k_{\parallel}$ is the same as
the cuts in Fig. \ref{Mapping2D}. Intensity maps in Figs.
\ref{Mapping3D}(a) and \ref{Mapping3D}(b) illustrate the
cross-sections of the Fermi surfaces by the $k_{\parallel}$-$k_z$
plane around the center and the corner of the 2D BZ, respectively.
Note that the intensity asymmetry observed with respect to the
$k_{\parallel}$=0 line is due to photoemission matrix elements. By
symmetrizing the plots in (a) and (b), we have obtained the
experimental FSs as shown in panel (c). For comparison, results of
band-structure calculation for $x$=0.4 and 0.6 are shown in panels
(d) and (e), respectively. Here, the band-structure calculation
has been performed for BaFe$_2$As$_2$ with the experimental
lattice constants for BaFe$_2$(As$_{1-x}$P$_x$)$_2$ with $x$= 0.4
and 0.6 \cite{Kasahara} in order to look into the effects of the
pnictogen height. As illustrated in the figure, we denote the hole
FSs around the center of the 2D BZ by $\alpha$, $\beta$, and
$\gamma$, respectively, and the outer and inner electron FSs
around the corner of the 2D BZ are denoted by $\delta$ and
$\epsilon$, respectively. Note that the $\beta$ and $\gamma$ FSs
intersect each other \cite{SO}.

As shown in Fig. \ref{Mapping3D}(c), the inner electron FS
$\epsilon$ exhibits warping qualitatively consistent with the
calculation. Correspondences between the observed hole FSs and the
calculation are not straightforward because we have observed only
two hole FSs as shown in Fig.\ref{Mapping2D} and Fig.
\ref{Mapping3D}(a) while the band-structure calculation predicts
three hole FSs. As for the inner hole FS around the $\Gamma$
point, a very recent ARPES result on the similar system
EuFe$_2$(As$_{1-x}$P$_x$)$_2$ has revealed that this FS has
$d_{xz/yz}$ orbital character \cite{Thirupathaiah}. Also, the
matrix element of the $d_{xy}$ orbital around the $\Gamma$ point
should be much smaller than those of the $d_{xz/yz}$ orbital
\cite{Zhang}. Therefore, this FS can not be the $\alpha$ FS which
has nearly pure $d_{xy}$ orbital character but the
three-dimensional $\gamma$ FS with $d_{xz/yz}$ orbital character.
As for the observed two-dimensional hole FS, if we assume that
this FS is the $\alpha$ FS, the energy level of the $d_{xy}$ band
should be much higher than that of the $d_{xz/yz}$ band at the
$\Gamma$ point, which contradicts with the general prediction of
band-structure calculation. Therefore, the observed
two-dimensional hole FS is assigned to the $\beta$ FS of
$d_{xz/yz}$ orbital character. In Fig. \ref{Mapping3D} (c), we
illustrate the $\beta$ and $\gamma$ FSs according to our
observation. We could not identify the $\alpha$ FS probably
because the spectral intensity is too weak and/or the band is
nearly degenerate with the $\beta$ and $\gamma$ bands.

\begin{figure}
\includegraphics[width=7.8cm]{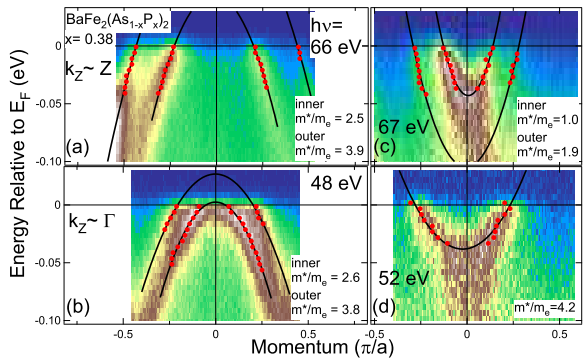}
\caption{\label{Mass}(Color online) Band dispersions around the BZ
center [(a)(b)] and the BZ corner [(c) and (d)] corresponding to
the cuts in Figs.\ref{Mapping3D} (a) and (b). Filled circles
indicate peak positions of the momentum distribution curves (MDCs)
obtained by taking second derivatives. Mass renormalization for
each band is obtained by fitting to parabolic dispersions.}
\end{figure}

In the dHvA measurements, the effective mass of the $\delta$
electron FS increases with decreasing $x$ down to $x\sim$ 0.4,
similar to the quantum critical behavior of heavy Fermion systems
\cite{Shishido}. The present ARPES data enable us to directly
determine the effective masses for each FS, by fitting a parabolic
band to the band dispersion as shown in Fig. \ref{Mass}. The
effective masses determined by ARPES as well as those derived from
the band-structure calculation are summarized in Table
\ref{masstable}. For the $\delta$ and $\epsilon$ FSs, the
effective mass ratios $m^*/m_e$, where $m_e$ is the free electron
mass, are estimated to be 2.8 and 2.0, respectively. Particularly,
the value for the $\delta$ FS is nearly the same as that of the
dHvA result for $x$=0.41 ($m^*/m_e\sim$3.3). Mass enhancement
factor $m^*/m_b$, where $m_b$ is the band mass obtained by the
band-structure calculation, varies from 1.7-3.9 for these FSs. As
a whole, these values are larger than those for the end members,
SrFe$_2$P$_2$ ($m^*/m_b\sim$ 1.3-2.1) and CaFe$_2$P$_2$,
($m^*/m_b\sim$ 1.5) \cite{SrdHvA,CadHvA}. Thus, the observed mass
enhancement factors indicate moderately strong electron
correlation effects and may be related to the proximity to the
quantum critical point.

\begin{table}[htbp]
\caption{Fermi surface volumes and effective masses of
BaFe$_2$(As$_{1-x}$P$_x$)$_2$ ($x$=0.38) determined by ARPES.
$m_b/m_e$'s are obtained by band-structure calculation. The
two-dimensional FS areas and the three-dimensional FS volumes are
expressed as a percent of the area of the 2D BZ and the volume of
the 3D BZ, respectively. For the hole FSs ($\beta$ and $\gamma$),
$m^*/m_e$'s determined from Fig.\ref{Mass} are listed because the
hole FSs are nearly isotropic in the $k_X$-$k_Y$ plane.
$m^*/m_e$'s for the anisotropic FSs ($\delta$, $\epsilon$) are
determined assuming $m^*=\sqrt{m^*_x m^*_y}$, where $m^*_x$ and
$m^*_y$ are masses in the two orthogonal axes. }
  \label{masstable}
\begin{tabular}{lclccccc}
        \hline\hline
FS & 3D volume & $k_z$ &2D area & $m^*/m_e$ & $m_b/m_e$ & $m^*/m_b$ \\
\hline
$\beta$ & 3.9 & $\Gamma$ & 3.9 & 3.8 & 1.3 & 2.9  \\
        &     & $Z$      & 3.8 & 2.5 & 1.2  & 2.1  \\
$\gamma$& 6.0 & $\Gamma$ & 1.0 & 2.6 & 0.9 & 2.9 \\
        &     & $Z$      & 16.3 & 3.9 & 2.3 & 1.7 \\
$\delta$& 5.3 & $X$      & 5.3 & 2.8 & 0.71 & 3.9  \\
$\epsilon$& 3.4& $X$     & 3.0 & 2.0 & 0.93 & 2.1 \\
        \hline\hline
    \end{tabular}
\end{table}

We have determined the three-dimensional volumes of the FSs as
listed in Table \ref{masstable}. To estimate the FS volume, we
have taken into account the warping of the FSs along the $k_z$
direction. The volume of the electron FSs $\delta$ (5.3 \% of BZ)
and $\epsilon$ (3.4 \% of BZ) are in good agreement with those for
$x$=0.4 obtained by the dHvA measurements, $\sim$6\% and
$\sim$3\%, respectively \cite{Shishido}. The experimentally
determined electron FSs are much smaller than those predicted by
the band-structure calculation, as clearly seen in Figs.
\ref{Mapping3D}(c)-(e). As for the hole FSs, the $\gamma$ FS shows
strong shrinkage around the $\Gamma$ point compared to the
calculation, while that around the $Z$ point has almost the same
diameter as the theoretical prediction. The total electron and
hole count from the observed FSs yields holes of 1.2 \% of the BZ
volume, indicating nearly compensated carriers although
contribution from the $\alpha$ hole FS is not included. Also, from
the effective masses $m^*$ in Table \ref{masstable} excluding the
$\alpha$ FS, the electronic specific heat coefficient $\gamma$ is
calculated to be $\gamma\sim$17 mJ/mol K$^2$, which is close to
$\gamma$= 16 mJ/mol K$^2$ estimated from specific heat measurement
\cite{Kim}. These results may be interpreted with the scenario
that the $\alpha$ band is pushed down below $E_F$ due to the
reduction of the pnictogen height as theoretically predicted for
BaFe$_2$P$_2$. To prove or disprove this possibility, further
investigation is necessary to detect the $d_{xy}$ band.

\begin{figure}
\includegraphics[width=8.3cm]{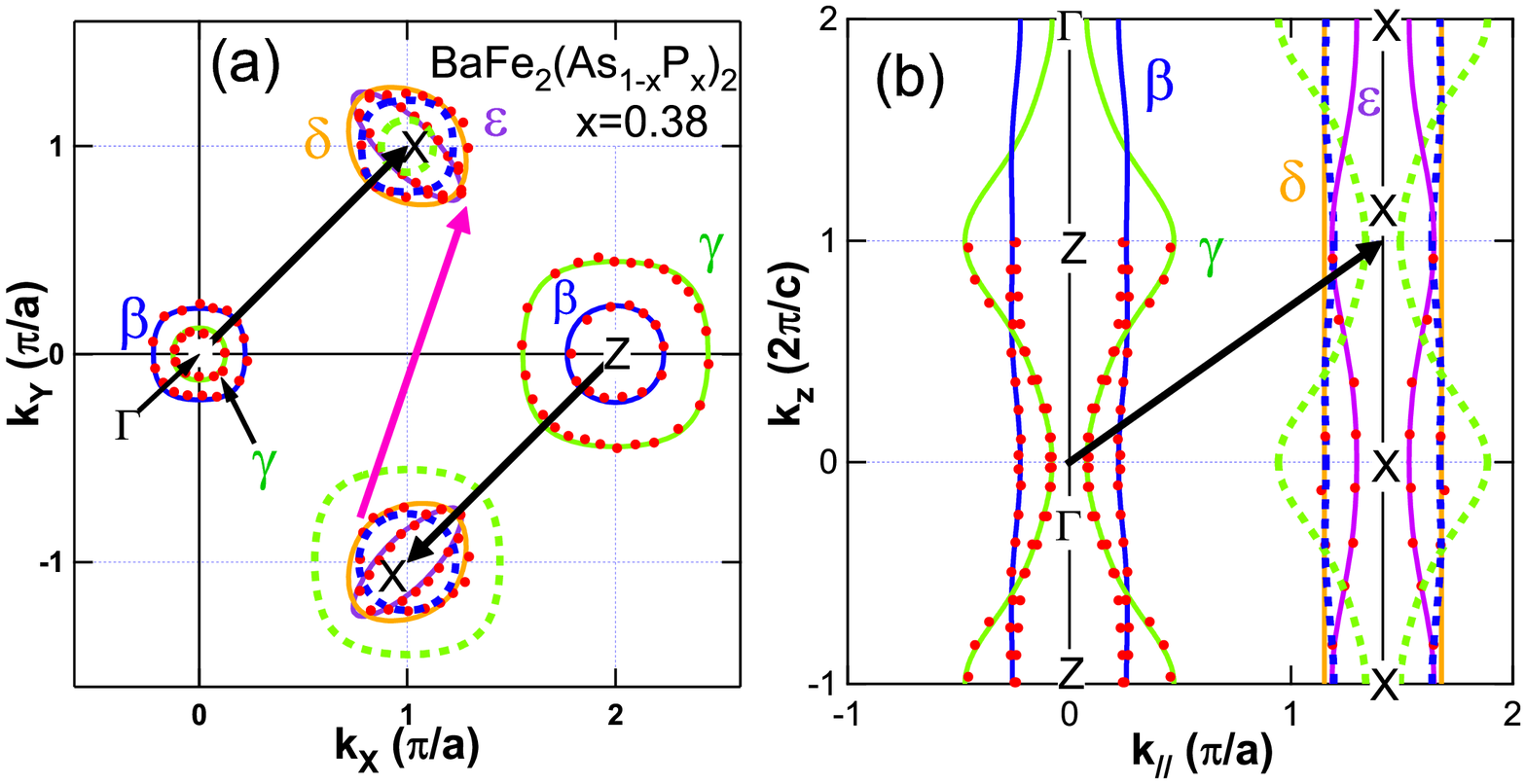}
\caption{\label{Nesting}(Color online) FSs determined by ARPES in
$k_X-k_Y$ plane (a) and $k_z-k_{\parallel}$ plane (b). Dotted
lines are hole FSs shifted by the antifferomagnetic vector (black
arrows). Partial nesting between the neighboring $\delta$ FSs is
indicated by a pink arrow.}
\end{figure}

Since FS nesting between electron and hole FSs has been discussed
as a necessary ingredient for the superconductivity in the
previous studies \cite{Ding, Terashima}, we shall discuss the
nesting properties of the FSs in three-dimensional momentum space.
The shapes of the observed FSs are reproduced in Figs.
\ref{Nesting}(a) and \ref{Nesting}(b). Here, hole FSs shifted by
the antiferromagnetic wave vector $(\pi/a,\pi/a,2\pi/c)$ of
BaFe$_2$As$_2$ \cite{Huang} are overlaid as dashed curves in Fig.
\ref{Nesting}(b). In a similar manner, the hole FSs are shifted by
an in-plane vector $(\pi/a,\pi/a,0)$ in Fig. \ref{Nesting}(a).
Note that the shifts by the both vectors are equivalent to test
the nesting conditions because the interval between two adjacent
$X$ points in the $k_z$ direction is $2\pi/c$.

According to the spin-fluctuation mechanism of superconductivity
\cite{Kuroki,Ikeda}, when a hole FS which has $d_{xy}$ orbital
character becomes absent, the fully gapped $s\pm$-wave
superconducting state becomes unstable and the nodal $s$-wave or
$d$-wave superconductivity is realized. This is because, in the
spin susceptibility, the structure which arises from the partial
nesting between the neighboring electron pockets of $d_{xy}$
orbital becomes dominant. As shown in Fig. \ref{Nesting}, the size
of the $\beta$ hole FS is nearly the same as that of the $\delta$
electron FS and both FSs have nearly cylindrical shapes, implying
good nesting. However, these FSs have different orbital character:
$d_{xz/yz}$ for the $\beta$ FS and $d_{xy}$ for the $\delta$ FS.
Hence, the $\beta$-$\delta$ nesting may have small contribution to
the spin susceptibility. The $\gamma$ FS shows strong warping and
therefore its nesting with the electron FSs is poor. Furthermore,
the $\gamma$ FS has $d_{3z^2-r^2}$ orbital character around the
$Z$ point, while $d_{3z^2-r^2}$ orbital character is almost absent
in the electron FSs. To summarize, the contribution to the spin
susceptibility from $\gamma$-$\epsilon$ nesting becomes small due
to the orbital character and the three-dimensionality of the
$\gamma$ FS.

On the other hand, partial nesting between the neighboring
$\delta$ electron FSs of $d_{xy}$ orbital character persists, as
indicated by the wave vector at the center of Fig. \ref{Nesting}
(a). Because the inter-band scattering between the electron and
hole FSs is reduced as mentioned above, the $\delta$-$\delta$
partial nesting may give a dominant contribution to the spin
susceptibility \cite{Kuroki} and may lead to the node in the
superconducting gap. That is, the three-dimensionality of the
$\gamma$ FS and the difference of the orbital character between
the hole and electron FSs suppress the fully gapped $s\pm$
pairing, leading to the nodal superconductivity.

An alternative scenario for the nodal superconductivity is an
appearance of horizontal nodes, which is likely to be realized in
the presence of warped hole FSs \cite{Graser}. In fact, in the
overdoped region of the electron-doped system
Ba(Fe$_{1-x}$Co$_x$)$_2$As$_2$, which has strongly
three-dimensional hole FSs \cite{Vilmercati,Malaeb}, it has been
pointed out that nodes occur in the Fermi surface that dominate
$c$-axis conduction \cite{Reid}. Possibly, the nodes result from
partial nesting within the $\gamma$ FS. If this scenario can be
applied to the present system, the strongly three-dimensional
$\gamma$ FS may have horizontal nodes. This should be clarified in
future studies.

In conclusion, we have experimentally determined the Fermi
surfaces of BaFe$_2$(As$_{1-x}$P$_x$)$_2$ ($x$= 0.38). We find
that the $\gamma$ hole FS has highly three-dimensional shape,
while the $\beta$ hole FS is nearly two-dimensional. Mass
enhancement for each band is stronger than those for the end
members, possibly due to enhanced electron correlation effect near
the quantum critical point. We have discussed nesting conditions
for the observed FSs. While the $\beta$-$\delta$ nesting looks
strong, the difference in their orbital characters may weaken the
inter-band scattering. The three-dimensionality and the orbital
character of the $\gamma$ FS also weakens nesting-induced
inter-band scattering. As a result of the poor nesting between the
electron and hole FSs, the partial nesting between the $\delta$
FSs and/or that within the $\gamma$ FSs become dominant and thus
the nodes in the superconducting gap are likely to be realized.
The present results give strong constraint on the theory of paring
mechanism and imply the importance of the three-dimensional FSs in
the nodal superconductivity.

We are grateful to K. Nakamura, T. Shimojima, W. Malaeb, K.
Kuroki, and S. Shin for informative discussions. This work was
supported by the Japan-China-Korea A3 Foresight Program from the
Japan Society for the Promotion of Science. Experiment at Photon
Factory was approved by the Photon Factory Program Advisory
Committee (Proposal No. 2009S2-005).

\bibliography{BaFeAsP}

\end{document}